# Mid-Air Haptics in Aviation - creating the sensation of touch where there is nothing but thin air

| | |
|---|---|
| **Alex Girdler** | **Orestis Georgiou** |
| **Collins Aerospace** | **Ultrahaptics** |
| **Sussex, UK** | **Bristol, UK** |
| alex.girdler@collins.com | orestis.georgiou@ultrahaptics.com |

## ABSTRACT

The exciting new technology known as mid-air haptics has been adopted by several industries including Automotive and Entertainment, however it has yet to emerge in simulated pilot training or in real-life flight decks. Full-flight simulators are expensive to manufacture, maintain and operate. Not only that, each simulator is limited to one aircraft type, which is inefficient for the majority of airlines that have several in service.

With the growing trend in touch-screen instrumentation, cockpit displays require the pilot's attention to be drawn away from their view out of the window. But by using gesture recognition interfaces combined with mid-air haptic feedback, we can mitigate this shortcoming while also adding another dimension to the existing technology for pilots already familiar with using legacy cockpits, complete with traditional instrumentation. Meanwhile, simulation environments using augmented and virtual reality technology offers quality immersive training to the extent that pilots can go from hundreds of hours of simulated training to being responsible for hundreds of lives on their very first flight. The software re-programmability and dynamic richness afforded by mid-air haptic technologies combined with a basic full-motion platform could allow for an interchange of instrumentation layouts thus enhancing simulation immersiveness and environments. Finally, by borrowing and exploring concepts within the automotive sector, this concept paper presents how flight deck design could evolve by adopting this technology. If pilot testimony suggests that they can adapt to virtual objects, can this replace physical controls?

## ABOUT THE AUTHORS

**Alex Girdler** spent most of his childhood sketching vehicle design concepts and was fortunate enough to go on to spend four years studying for a BA (Hons) degree in the subject at The University of Huddersfield. During the sandwich course, he fell in love with computer-aided design and modeling. Upon graduation, he worked on a fixed-term contract for a digital agency on a vehicle modeling project. Since then, he enjoyed a decade as an Environment Artist for Collins Aerospace, working on projects for some of the biggest names in the commercial aviation industry. In 2018 he was promoted to the position of Regional Lead, looking after 200+ models in the Americas and Asia catalogue. His strengths in his current role lie very much in bringing together products, software tools and emerging modeling techniques for an improved product.

**Orestis Georgiou** has a PhD in Applied Mathematics from the University of Bristol. He has published over 65 articles in leading journals and conferences of Mathematics, Physics, Computer Science, Engineering and Medicine, two of which received Best Paper Awards. Orestis is passionate about innovation, impact, and collaboration, and is currently a Marie Curie Fellow at the IRIDA research centre for communications and Director of Research at Ultrahaptics where he is co-PI of the FET Open projects Levitate and H-Reality. Orestis was recently awarded the 2019 IEEE Heinrich Hertz Award for his contributions towards enlarging the field of communications engineering through the mathematical analysis of LoRa networks.

# Mid-Air Haptics in Aviation

| Alex Girdler | Orestis Georgiou |
|:---:|:---:|
| **Collins Aerospace** | **Ultrahaptics** |
| **Sussex, UK** | **Bristol, UK** |
| **alex.girdler@collins.com** | **orestis.georgiou@ultrahaptics.com** |

## INTRODUCTION

An entire industry exists to install real-life flight decks, display and visuals systems in a device that replicates flight for pilot training. Since the age of model boards, where a pilot controlled the movements of a camera over a physical terrain model, the aim has been to reduce the associated costs and risks with training using real-life flight. This training paradigm can be taken to the next level using existing technology.

Firstly, here are the drawbacks of the current commercial simulation model:

- Certified flight training devices (FTD) (no motion) cost hundreds of thousands of dollars.
- Certified full flight simulators (FFS) (full motion) cost tens of millions of dollars.
    - Each FFS is limited to one aircraft type, and the majority of airlines have several in service.
    - From an environmental perspective, each FFS has a huge carbon footprint.

Ultrasonic mid-air haptics is an emerging new technology that enables users to touch and feel virtual 3D holograms such as buttons and dials with their bare hands, without having to wear or hold any specialised controllers. This technology can potentially be harnessed to replicate the physicality of virtual controls of multiple aircraft, and be interchangeable (software programmable) at the touch of a button.

With a basic motion platform, pilot seating and virtual or augmented reality (VR/AR) display offerings, pilots could train for multiple aircraft, all on this single (yet very versatile) setting. They would *see* the controls through the AR/VR headset, they would *see* the synthetic environment, they would *feel* full-motion, they would *hear* common sounds such as the landing gear deployment and, finally, they would *feel* the virtual controls which are visible to them; a multi-modal immersive environment.

## TRENDS

Are trends in simulation heading towards a more accessible device? The Federal Aviation Administration (FAA) amended part 61.57 of the Code of federal regulations (CFR), title 14 of the federal aviation regulations (FAR). In short, this allows pilots in command to maintain Instrument flight rules (IFR) currency with flight time logged on basic aviation training devices (ATDs), and potentially even on a simulator built at home. It is therefore an interesting and timely research topic to explore, develop, and validate new haptic interfaces and their use in aviation.

During the 2017 Interservice/Industry Training, Simulation and Education Conference (I/ITSEC), it was summarised that "lots of training flying will break the bank, therefore more simulation is needed" and that "we need to embrace simulation to a level not seen before."

## INTRODUCTION TO GESTURE INPUT CONTROL INTERFACES

Mid-air gestures and control interfaces are generally facilitated by optical tracking systems that apply machine vision algorithms to recognize hand gestures (HGs) that are translated into the input commands of human-machine interfaces (HMIs). Gesture based control interfaces have already become almost ubiquitous in our daily lives. We use HGs to interact with our appliances, smartphones, electronic devices and, more recently, virtual and augmented environments. The widespread availability of these solutions has been made possible by a new generation of tracking devices and gesture recognition techniques (Microsoft's Kinect, Intel's RealSense, Leap Motion etc.), which have progressively become more accurate, reliable, and affordable. These tracking systems have been embraced by the research and development communities and have in some cases morphed into bespoke tracking solutions that are increasingly being deployed in more and more demanding situations.

Already, car manufacturers such as BMW, Mercedes, VW, Cadillac, Jaguar, and Hyundai see great potential in mid-air HG interfaces and have integrated HG interactions into some of their premium models. A key advantage is their ability to reduce the mental and visual demand of the driver by making good use of the driver's peripheral vision and proprioceptive awareness related to the motion of their arms without having to look at the centre console HMI, thereby decreasing visual and mental demand, and reducing visual distraction from the road.

Touch-less controllers are also increasingly popular in the context of AR/VR applications. Microsoft's Hololens 1 and 2 allows wearers to control the operating system (OS) embedded in the AR system by hand tracking gestures such as those of pinching and tapping. Meanwhile Leap Motion have been active in both VR and AR realms by working with commercial VR headsets, e.g. those made by Oculus and HTC, and their own AR headset currently known as project North Star. Touch-less controllers and hand tracking technology expands the immersive quality of the user interface in AR/VR by enabling the user's hands to become part of the simulation without the additional wearable tech required. Importantly, this development overcomes cognition limitations over older generation training videos and controller-based simulations, and therefore allows the user to directly reach out with their hands into VR and engage with objects and scenes. This interactivity makes the overall experience more tangible and immersive, an important component when delivering training in simulation.

However, the immediateness and natural language interface of the hand (Bulwer, 2003) provided by touchless controllers has a significant drawback: it lacks the force and tactile feedback that comes from interacting with a physical device. Studies performed in several domains have highlighted the importance of haptic feedback and its derived benefits in accuracy (Robles-De-La-Torre, 2006), performance (Stone, 2000), embodiment (Kilteni, 2012), sense of agency, i.e., the subjective experience of voluntary control over your actions (Cornelio Martinez, 2007), immersiveness, and overall user experience (Danieau, 2012).

It is therefore paramount that HMIs find alternative ways of re-instilling physicality to virtual controls by other means. For instance, the haptic sensory modality can be augmented, supplemented, or even substituted by other modalities. The latter are often referred to as pseudo- or pre- haptics (Lécuyer, 2009) and are known to be quite effective since senses need not always accurately reproduce reality, since our imagination can often fill in the gaps. At the same time however, it is simplistic to say that adding haptics to HMIs directly translates to better raw performance. Indeed, most of the movements we execute are pre-planned; recall that the human brain is often called a "prediction machine". Therefore, if the sensory feedback we receive does not match our expectations it may cause a disconnect. Hence, an important role of providing high-quality haptics in HMIs is to satisfy the sensory expectations of the users or at least to deliver consistent and reliable feedback (MacLean, 2000).

**INTRODUCTION TO MID-AIR HAPTIC FEEDBACK TECHNOLOGY**

The latest developments in mid-air haptic technology has made it possible for designers to build touch-less interfaces that also incorporate tactile feedback. Such technologies include, for example, mid-air focused ultrasonic haptic interfaces (Carter, 2013), air jets (Sodhi, 2013), air cannon (Gupta, 2013), and femto-lasers (Ochiai, 2016), each of which is suitable depending on the desired application. With the exception of ultrasound mid-air haptics, these technologies currently exist only in research laboratories and have limited range (<1m) and resolution.

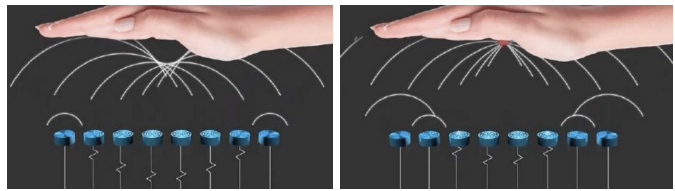

**Figure 1: Multiple ultrasonic waves converging to a point to create a tactile sensation in mid-air.**

Ultrasound mid-air haptics technology uses electrical signals that drive a set of ultrasonic speakers (or transducers) such that ultrasound waves interfere constructively at one or more focus points in space, such that a tactile sensation is felt when touched by the hand of a user (see Figure 1). The haptic sensation is a vibrotactile one, not a force feedback one. Hence it is most closely associated with touch sensation felt by the skin, rather than those felt by our muscles and proprioceptive system.

This touchless technology was first demonstrated in Japan (Hoshi, 2010) and commercialized by the UK-based company Ultrahaptics, who build and sell ultrasound haptic devices also called phased arrays. When a user touches a virtual object, the contact points between their bare hand and the object are recorded and one or more ultrasonic focus points can be made to "jump" from point to point (in some predefined order and speed) such that the object surface is felt by the intersecting palm or fingers thus creating a tactile hologram.

Ongoing research and development is targeted towards optimising, customising, and personalising haptic rendering algorithms for the best possible subjective user perception of haptic shapes and their sensation. Moreover, these efforts do not only target haptic algorithmic designs but also expand upwards towards applications and use cases in automotive (Harrington, 2018) (see Figure 2) and digital signage (Georgiou, 2019) (Corenthy, 2018), thereby forming close links with user interface and interaction design. Similarly, there is an active research drive towards the fabrication of smaller and cheaper ultrasonic transducers (van Neer, 2018) (Ito, 2016). Currently, most stable demonstrations of ultrasonic mid-air haptics operate at 40 kHz and use the Murata MA40S4S piezoelectric transducers which operate at this frequency.

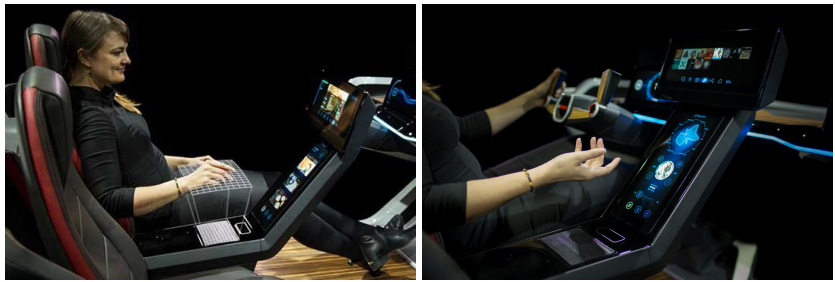

**Figure 2: Concept user interface presented by Bosch and Ultrahaptics at the CES 2017.**

Existing development kits and products offered by Ultrahaptics utilise 256 small MA40S4S transducers that together with the appropriate driver electronics form a phased array capable of steering and focusing the ultrasonic energy. For example, the Ultrahaptics STRATOS Inspire (USI) platform shown in Figure 3 can project haptic sensations at a distance of up to 70 cm from the array and within a cone-angle of 90 degrees. This volume is analogous to the hand tracking volume afforded by the Leap Motion sensor integrated within the USI. The USI consumes approximately 80 watts of electrical power at its peak. Adding transducers increases the maximum effective range, while having multiple panels of transducer arrays side by side or at different orientation angles increases the haptic interaction region. This is particularly useful in simulation environments where the virtual control panel extends across a large field of view. Within a virtual environment, e.g., in VR or on a screen, the user's hand can be replaced by a digital avatar that triggers the mid-air haptic effects whenever it collides with a virtual object (see Figure 3).

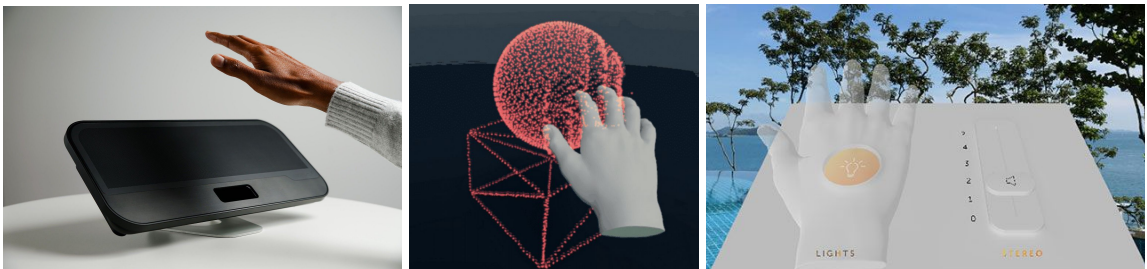

**Figure 3: Left: Typical mid-air haptic interaction with a USI device. Middle: Example interaction of a hand avatar with a particle point-cloud of a sphere. Right: Example of a hand avatar pressing a virtual button.**

There are currently two ways of inducing a tactile sensation using focused ultrasound. The simplest way involves switching the ultrasound waves off and on again at a rate of between 10 and 400 times per second (10Hz - 400Hz), in order to stimulate the vibration-sensitive receptors in the users' skin as they hover them over the device in midair. This method is called amplitude modulation (or AM) and is very effective in creating localised tactile sensations for simpler applications. The second method moves the focal points away from the user's hand and back again (instead of turning them on and off) at a frequency that is perceptible by touch receptors. Hence, this technique is called spatio-temporal modulation (STM) – the ultrasonic waves are modulated both in space and time. In the same way as touch receptors in users' hands can detect the on-off switching of ultrasound waves, they can detect the change in sound pressure as the focal point moves away from and back towards their hand. Together, AM and STM form a spectrum of tactile sensations that can be rendered on a user's palm and fingers to create virtual objects and shapes thus re-instilling physicality and the sense of agency to a plethora of applications in augmented and virtual reality.

To facilitate application development of ultrasonic haptic technology and to move it up the technology readiness levels (TRL), a software development kit (SDK) has been assembled through which commands in C++ or C# programming languages can be translated into AM and STM ultrasound tactile sensations. Already, there are template code snippets (aka blocks) for a variety of haptic sensations and 3D virtual objects that can be easily ported into 3D game engines like Unity and Unreal (where many AR/VR applications are usually designed), thus *haptifying* and enriching existing audio-visual interactive environments. Importantly, a number of flight simulators are currently built in either of these two 3D game engines. The haptic sensation templates are encompassed within the so called sensation core library (SCL) and are constantly updated.

Finally, the computations needed for the haptics SDK have been embedded in ready-to-use hardware such as field-programmable gate arrays (FPGAs) that drive the electronics behind these phased transducer arrays. The carrier frequency is 40 kHz and each pressure point can move in space freely (resolution of 4 mm in a 120° field of view) and extremely rapidly (its position in space can be updated at a rate of up to 40 kHz) to render 3D objects (e.g., cubes, buttons and dials) that a user can touch and feel.

## VIRTUAL, AUGMENTED, AND MIXED REALITIES

In recent years there has been an explosion of interest in head mounted displays (HMDs) and the use of advanced graphics and machine vision to enhance or extend the way we experience real and virtual worlds. The scientific and industrial communities for these technologies both have a strong history, going back to Sutherland's ultimate display in 1965 (Sutherland, 1965) and Furness Super cockpit in 1986 (Furness III, 1986) to name a few, have in recent years joined forces therefore pushing even harder for the mass deployment of such systems. We briefly simplify the current taxonomy of realities available today before we discuss their application and opportunities in aviation simulators.

**Virtual Reality:** The ultimate goal of VR is true immersion, described as  the feeling of embodiment into another place or space without physically actually being there. This sort of ability has therefore strong applicability to the (serious) game industry. It is also a hugely versatile tool for the psychological exploration of the human condition, its psychoanalysis and therefore rehabilitation (Rothbaum, 2001). In some sense, VR is therefore a hardware innovation, not restricted to HMDs since it can include suits, gloves, custom controllers, etc., and is comparable to the invention of the mouse and keyboard that analogously offer more intuitive control over, and deeper level immersion into, digital experiences.

**Augmented Reality:** The technology to overlay information on top of real-world images and video is well established, but the key differentiation of AR is how it can extract and blend this information to create a new reality. As a result, AR is a software innovation. It relies heavily on machine learning and computer vision rather than hardware and is not limited to head mounted displays. In fact with the recent release of Apple's ARKit , AR has enjoyed mass adoption and application development on iOS mobile devices such as iPads and iPhones.

**Mixed and eXtended Reality (XR):** These are more abstract concepts than AR and VR that somehow combine the two hardware and software innovations to offer application specific solutions and therefore can adhere to higher complexity requirements. The general understanding at the time of writing is that MR is a combination of AR and VR with a focus on attaining realism, immersiveness, and productivity, while XR is an all-encompassing term that includes AR, VR and MR with a strong focus on enhancing the experienced reality.

## MIXED REALITY TECHNOLOGY FOR SIMULATION AND TRAINING

The aviation simulation and training industry is constantly challenging and upgrading its systems. One of the current debates relates to the effectiveness of training pilots using MR systems. Some argue that no current solution can replace the tried-and-tested flight simulators that provide pilots with realistic simulation environments of aircraft cockpits where they can, for example, reach out for their oxygen mask when the cockpit fills with smoke or feel the subtle vibrations of the plane due to air turbulence through a real cockpit seat. However, rapid progress is being made by MR that presents a cheaper and more flexible alternative to high fidelity physical simulators. Already, MR simulators are being used as an alternative to video tutorials by the U.S. Air Force and private flight schools. How far away are MR simulation systems from completely replacing expensive and clunky physical simulators?

Within simulation training devices, mixed reality (MR) display products such as Collins Aerospace's Coalescence™ allow not only a synthetic environment (SE) to be viewed, but also the user's hands, props and real world view. This is achieved by three hardware subassemblies: an HMD, a video merge processor and SE-renderer complete with an SDK, thus allowing for frequent updates and modifications (see Figure 4).

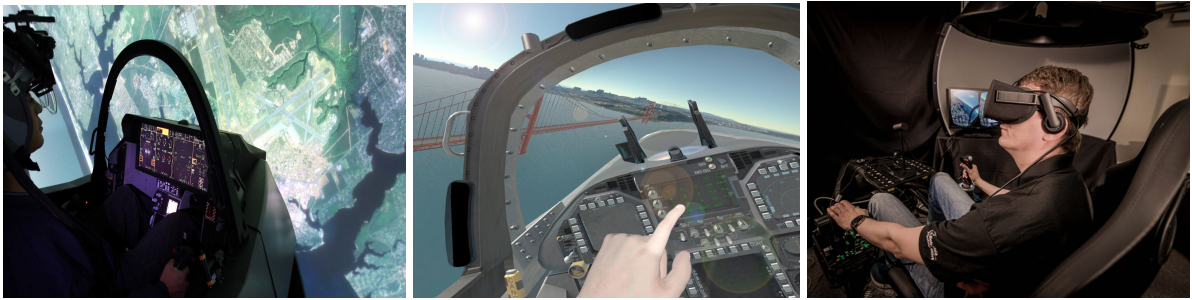

**Figure 4: Existing MR flight simulator concepts that exploit AR/VR HMDs and a SE. The simulator in  the middle figure also uses hand tracking technology for virtual interaction with control panels.**

The addition of mid-air haptics devices could bring a new dimension and versatility to this immersive training environment, which provides a seamless and enhanced MR.

**CONCEPT PROTOTYPE**

The use of mid-air haptics for buttons, switches and dials scattered throughout the cockpit would prove relatively simple to implement. Fly-by-wire cockpit controls however consist of side sticks, rudder controls, thrust controls, etc. thus making the concept prototype design non-trivial. Therefore, achieving the *look* (through AR/VR headsets) and *feel* (via mid-air haptics) of controls such as switches, dials and knobs could prove a strong first step in exploring whether controls can exist and function synthetically. As a second step, engine controls such as thrust lever/throttle, landing gear, and brakes could also be haptically rendered within the virtual simulation, either via mid-air haptics or by physical controls as those required in current Level A-D certified FFSs (see below).

**3D Control Panels with Hundreds of Buttons**

The Bombardier CRJ 200 aircraft has about 350 different knobs, switches, and other controls reachable by the two pilots. The Airbus A330 has over 200 buttons on the overhead panel alone, whilst the Boeing 737 cockpit has undergone dozens of iterations over the past 50 years. Physical flight simulators therefore struggle to emulate the zoo of possible cockpit configurations and hundreds of control buttons. Touch screen liquid crystal displays (LCDs) have certainly alleviated this task significantly as controls can now be resized, reshaped and replaced freely on the 2D screens. To achieve true, 3D placement and augmentation however, we propose the following concept setup:

Within the flight simulator, a user typically finds themselves sitting in the pilot seat surrounded by the synthetic (virtual and physical) MR cockpit of the virtual aircraft model. Furthermore, the aircraft may be stationary or in motion, hence a simulation of the dynamic outside environment is also visible through the flight deck windows. The user is able to reach out and interact with the cockpit control panel using both their hands and fingers, e.g., to flick switches on/off or move levers up/down. Already several such systems exist on the market such as the Collins Aerospace Coalescence and the Bohemia Interactive Simulations (BISim) shown in Figure 4 above. Unlike in existing systems however, our concept prototype ambitiously envisions the complete removal of all physical controls and props, and their replacement by ultrasonic mid-air haptic replicas, thus not sacrificing their tactile properties.

To that end, we propose using four USI mid-air haptics device and two Lighthouse positional tracking devices mounted on a metallic scaffolding bracket as seen in Figure 5 below. The Lighthouse trackers provide accurate positional information about the AR/VR HMD and any other trackable accessories. The user of the system sits on a chair that is approximately at arm's length from the mid-air haptic system. The chair could be placed within a full-motion platform or itself be force feedback and motion actuated. The four USI devices come pre-integrated with four Leap Motion sensors thereby providing accurate hand and finger tracking of the user's interactions in the relatively large interaction volume. The USI themselves are positioned and angled such that they provide sufficient haptic coverage of the interaction zone. This means that haptic feedback can be instantaneously generated anywhere within the highlighted green volume to emulate, e.g., the tactility of a button click, a moving lever or a flicking of a switch. Further, the four USI devices are pre-calibrated relative to the position of the VR Lighthouse and hence the scene coordinates within the 3D simulation software. To maintain synchronicity across the different components and also to minimize development complexity, the complete setup can run on a single PC with a powerful graphics card.

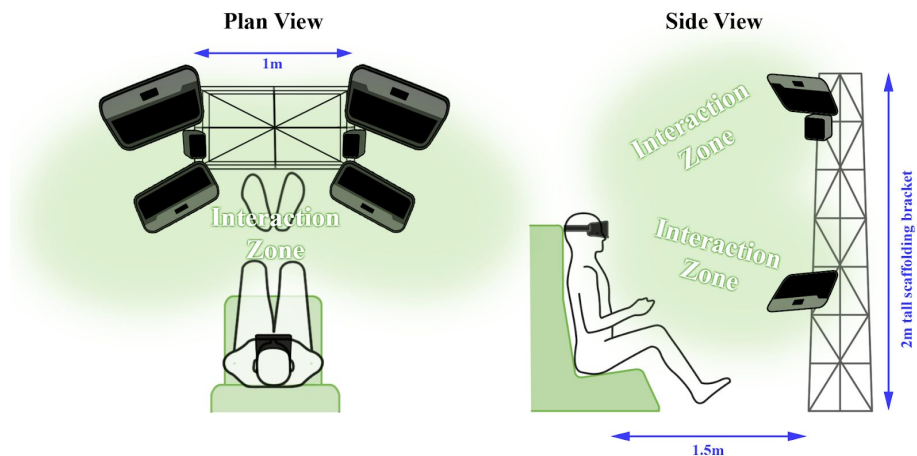

**Figure 5: Plan and Side view of the schematic setup of a low fidelity cockpit VR simulator.**

The software integration of ultrasonic mid-air haptics into the VR simulation graphics scene implicitly requires the haptic labeling of the different controls using a library database of haptic sensations. For example, and making direct reference the concept drawings of Figure 6 below showing a simplified A320 flightdeck, the rotary dials in panel (1) could be emulated using the "dial" sensation projected onto the users index and thumb fingers; the press-down buttons in panel (2) could be emulated using the "click" sensation projected onto the index finger; the landing gear lever in panel (3) could be emulated using the "presence" sensation projected onto the index, thumb and middle fingers; and finally the Throttle Control Module in panel (4) could be emulated by the "line" sensation projected onto the middle of the palm. These example mid-air haptic sensations mentioned here are already contained in the SCL and would only activate when the user is interacting with the specific control part. Once haptic labels are inserted into the simulation software for several virtual cockpits, the user can seamlessly switch training modes between different aircraft. For FTD Level 5 certification, additional physical controls could be incorporated for panels (3) and (4) thereby somewhat limiting the versatility of the simulation system. These controls however would not obstruct and hinder the mid-air haptification of non-primary flight controls found in panels (1) and (2). More advanced and custom mid-air haptic sensations could also be incorporated in future simulator iterations as required.

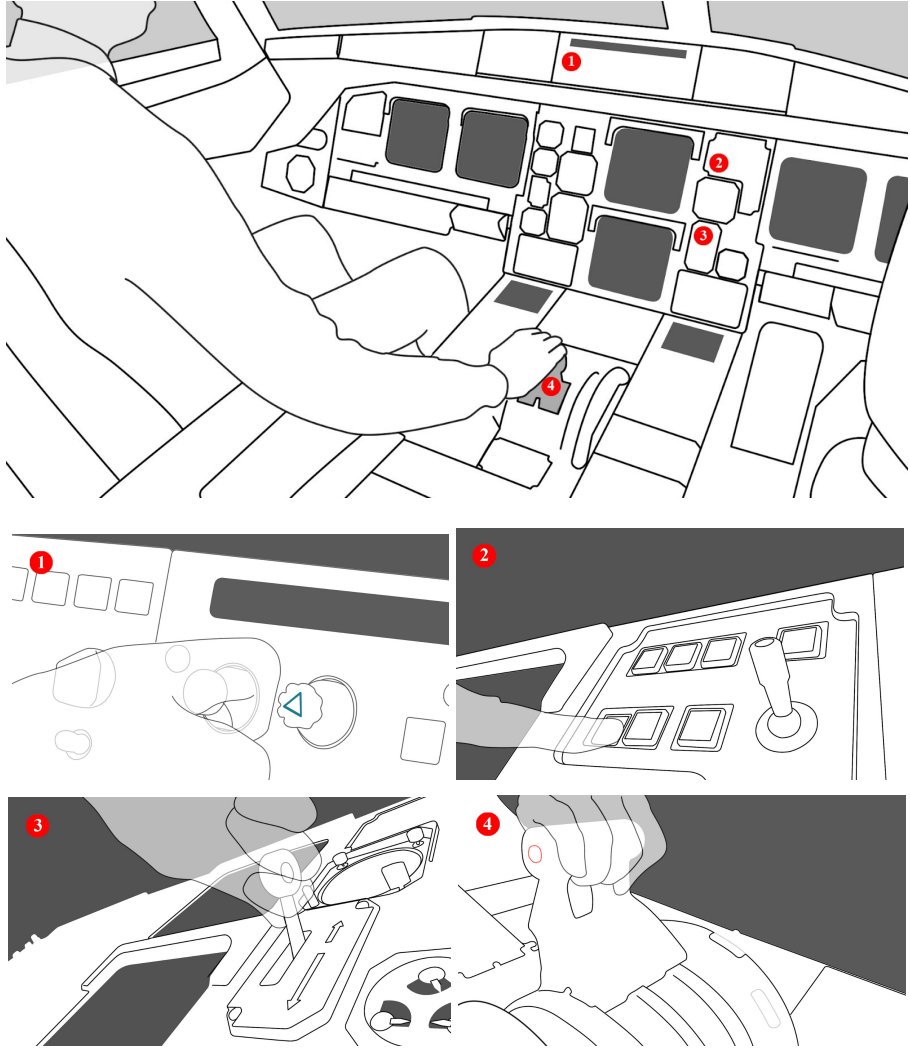

**Figure 6 - Concept drawings, taking the A320 flightdeck as an example. Here buttons, knobs, dials and switches are represented by virtual mid-air haptic objects.**

**Future Testing, Validation, and Optimisation of the Concept Prototype**

The quality of flight simulators strongly contributes towards the safety of aviation activities in general however they can often be influenced by multiple interacting factors. How humans interact with simulators plays an essential role in these factors and can often lead to poor flight simulation and training performance. Thus, the study of the relationship between human factors and flight simulator performance is paramount in order to test, validate, and optimise the adoption and effective use of new technologies in flight simulators and in aviation in general.

To that end, a necessary successor to the current concept paper is a sequence of human factors and ergonomics studies geared towards data collection that can help assess the use of mid-air haptics in aviation simulators, in general, and specifically with respect to pilot training. For example, these studies could include quantitative measures of flight control and reaction times, but also subjective feedback on the cognitive and mental workload associated within the training simulator, e.g., through completing NASA-TLX questionnaires. While such an undertaking is time consuming, several computational models exist that can be adapted and used to help evaluate the human factors manifested in this new human-machine interface such as any physical limitations and cognitive constraints on pilot behaviors and flight performance (Steelman, 2011).

**OUTLOOK FOR THE CHALLENGES AHEAD**

The below study discusses if certification standards (A, B, C or even D) can be achieved in full flight simulators (FFS) without projectors, without a cockpit and without instrumentation. It also looks at other simulation products, including lower-level flight training devices (FTD) and their (Level 4/5/6) certification standards.

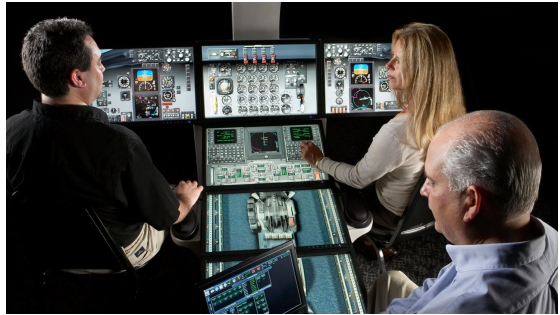

**Figure 7 - FAA approved Level 4/5/6 Flight Training Device (FTD)**

**Consideration of regulatory FTD criteria**

As Figure 7 above illustrates, the addition of mid-air haptics arguably would go beyond the scope of a FTD (no motion) device with their flat-panel instrumentation. The FAA descriptions state each level of FTD as follows:

**(1) Level 4:** "A device that may have an open airplane-specific flight deck area, or an enclosed airplane-specific flight deck and at least one operating system. Air/ground logic is required (no aerodynamic programming required). All displays may be flat/LCD panel representations or actual representations of displays in the aircraft. All controls, switches, and knobs may be touch sensitive activation (not capable of manual manipulation of the flight controls) or may physically replicate the aircraft in control operation." *No potential risks identified here, only opportunities.*

**(2) Level 5:** "A device that may have an open airplane-specific flight deck area, or an enclosed airplane-specific flight deck; generic aerodynamic programming; at least one operating system; and control loading that is representative of the simulated airplane only at an approach speed and configuration. All displays may be flat/LCD panel representations or actual representations of displays in the aircraft. Primary and secondary flight controls (e.g., rudder, aileron, elevator, flaps, spoilers/speed brakes, engine controls, landing gear, nosewheel steering, trim, brakes) must be physical controls. All other controls, switches, and knobs may be touch sensitive activation." *There is a risk here that the mid-air haptic solution cannot be classified as a true representation of physical controls.*

**(3) Level 6:** "A device that has an enclosed airplane-specific flight deck; airplane-specific aerodynamic programming; all applicable airplane systems operating; control loading that is representative of the simulated airplane throughout its ground and flight envelope; and significant sound representation. All displays may be flat/LCD panel representations or actual representations of displays in the aircraft, but all controls, switches, and knobs must physically replicate the aircraft in control operation." *There is a risk that the mid-air haptic solution cannot be classed as a true 'representation' due to virtual controls.*

**(4) Level 7:** "A Level 7 device is one that has an enclosed airplane-specific flight deck and aerodynamic program with all applicable airplane systems operating and control loading that is representative of the simulated airplane throughout its ground and flight envelope and significant sound representation. All displays may be flat/LCD panel representations or actual representations of displays in the aircraft, but all controls, switches, and knobs must physically replicate the aircraft in control operation. It also has a visual system that provides an out-of-the flight deck view, providing cross-flight deck viewing (for both pilots simultaneously) of a field-of-view of at least 180° horizontally and 40° vertically." *There is a risk here that the mid-air haptic solution cannot 'physically replicate the aircraft in control operation' due to virtual controls.*

**Consideration of regulatory FFS criteria**

Certification criteria from regulatory authorities states that every single audio and visual element of a pilot's simulation experience must exactly represent a real-life experience. Achieving this realism is vital: pilots can go from hundreds of hours of simulated training to being responsible for hundreds of lives on their very first flight.

Can mid-air haptic representations of physical controls, switches and knobs be certified for pilot training? Are the ultrasonic haptic representations of these controls good enough to merit certification?

Concentrating solely on the FAA regulatory authority for the purposes of this paper, they list extensive visual requirements for Class 1 airports through detailed functions and subjective tests. These relate chiefly to the real-time rendered visual scene, which traditionally is back-projected onto a collimated display through three projectors, giving a 180° field of view. This is a very complex and expensive setup. What is the likelihood that these criteria can be met when viewing through MR headsets? The FAA Qualification Performance Standards (QPS) are specified as "the minimum airport visual model content and functionality."

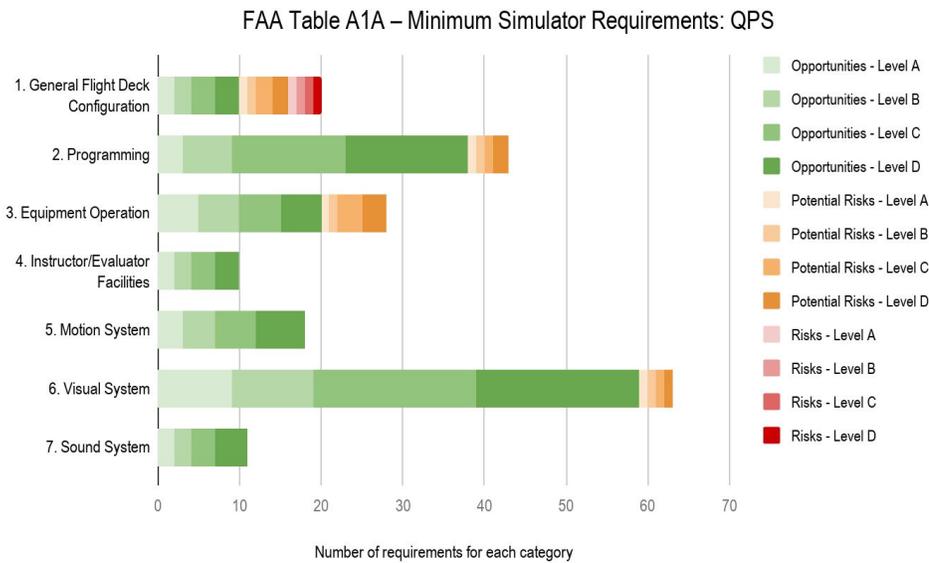

**Figure 8. Opportunities vs risks regarding FAA FFS Qualification Performance Standards (QPS)**

The chart in Figure 8 summarises which FFS simulator level would be unaffected by such a training environment. Of the hundreds of criteria listed, the opportunities as displayed as green, potential risks in amber and definite risks to meeting regulatory requirements are in red. Criteria applicable to each simulator level are displayed within each as a shade of colour. Details of risks and potential risks (the reds and ambers above) to follow:

**1. General Flight Deck Configuration:**
- "The simulator must have a flight deck that is a replica of the airplane simulated with controls, equipment, observable flight deck indicators, circuit breakers, and bulkheads properly located, functionally accurate and replicating the airplane'. The appearance of the simulated instrument, when viewed from the principle operator's angle, should replicate that of the actual airplane instrument." *Applicable to levels A,B,C and D, there is a risk that the mid-air haptic solution cannot be classed as a true 'replica' due to virtual controls.*

**2. Programming:**
- "Relative responses of the motion system, visual system, and flight deck instruments, measured by latency tests or transport delay tests. Motion onset should occur before the start of the visual scene change (the start of the scan of the first video field containing different information) but must occur before the end of the scan of that video field. Instrument response may not occur prior to motion onset. Test results must be within the following limits:

- ○ "300 milliseconds of the airplane response." *Applicable to levels A and B, there is a potential risk here that latency may be inherent. Testing this is outside of the scope of this concept paper.*
- ○ "100 milliseconds of the airplane response (motion and instrument cues). 120 milliseconds of the airplane response (visual system cues)." *Applicable to levels C and D, a potential risk exists without thorough latency testing.*

- "The aerodynamic modeling in the simulator must include:

  - ○ Normal and reverse dynamic thrust effect on control surfaces." *Applicable to level D, there is a potential risk here if the developed concept does not create the necessary feedback.*

**3. Equipment Operation:**
- "The simulator must provide pilot controls with control forces and control travel that correspond to the simulated airplane." *Applicable to levels A, B, C and D, the travel and forces of the virtual objects would have to be precise in replicating the physical counterparts.*
- "Simulator control feel dynamics must replicate the airplane. This must be determined by comparing a recording of the control feel dynamics of the simulator to airplane measurements. For initial and upgrade qualification evaluations, the control dynamic characteristics must be measured and recorded directly from the flight deck controls, and must be accomplished in takeoff, cruise, and landing flight conditions and configurations." *Applicable to levels C and D, the dynamics of the virtual objects would have to be precise in replicating the physical counterparts.*
- "For aircraft equipped with a stick pusher system, control forces, displacement, and surface position must correspond to that of the airplane being simulated. A Statement of Compliance (SOC) is required verifying that the stick pusher system has been modeled, programmed, and validated using the aircraft manufacturer's design data or other acceptable data source. The SOC must address, at a minimum, stick pusher activation and cancellation logic as well as system dynamics, control displacement and forces as a result of the stick pusher activation. Tests required." *Applicable to levels C and D, virtual controls would have to achieve this SOC.*

**4. Instructor or Evaluator Facilities:** *No concerns as the proposed solution does not extend to the Instructor or Evaluator facilities, therefore a standard setup would be utilised.*

**5. Motion System:** *No concerns as a standard motion platform would be utilised.*

**6. Visual System:**
- "The visual system must be free from optical discontinuities." *Applicable to levels A,B,C and D, any visual anomalies within the proposed solution would not be certifiable.*

**7. Sound System:** *No concerns as a standard audio system would be utilised.*

**SUMMARY**

There is a growing demand for flight simulators in recent years due to the increase of air traffic and the increased acceptability of virtual training. There is, therefore, a great need for effective and affordable flight simulators that utilise new and emerging technologies. To that end, this paper has investigated the use of mid-air haptic technology in simulation. We have thus provided a brief overview of current trends including the use of AR/VR headsets for the visual simulation of different cockpits, the use of physical controls to re-create a synthetic MR environment, and finally the use of hand tracking technology to supplement or even replace the physical controls completely while maintaining tactility through the use of ultrasonic mid-air haptics. A whole section was dedicated to describing this enabling technology, and then detailing a concept prototype VR simulator. It was argued that by adding mid-air haptics to low-level FTDs with only their flat-panel displays acting as flight controls, a new dimension of immersion can be achieved. Moreover, in view of these touchless (non-physical) haptic devices, there is a possibility that an entirely new level of certification may be required.

When considering FFS devices, the addition of virtual objects through mid-air haptics allows for cuts in costs, a more accessible training solution and even the flexibility in aircraft type. The risks in achieving the highest FAA level D certification have been listed, and the development of a solution must work to progressively overcome them. The ultimate challenge, however, lies in demonstrating to the industry how this display technology has the capability to simplify training with its lack of physical props and real-world controls. To that end, prototype development and a sequence of testing and validation studies was proposed and discussed.

Since its inception, trends in pilot training have suggested that it is about creating the most realistic scenario in the most accessible way. There is clearly a gap to be filled, using this existing, cutting-edge mid-air haptic technology. If the automotive sector is already doing it, then why can't aviation?

Looking to real-life flight and aviation as a whole, as flat-panel displays become more and more commonplace, physical controls could be re-introduced in a virtual manner. Legacy pilots would then gain the tactile sensations that they learned to fly with. Flight decks could be customised to seek the needs of an individual pilot.

**ASSUMPTIONS**

- Replicating hard or solid surfaces through mid-air haptics requires advancements in programming and extra grunt and resolution from the hardware array of ultrasonic speakers. We assume that the technology will be able to meet these requirements as it matures. Alternative force feedback devices could be also be included.
- The ultrasound used for mid-air haptic feedback does not interfere with any other FTD/FFS components.

**ACKNOWLEDGEMENTS**

This project has received funding from the European Union's Horizon 2020 research and innovation programme under grant agreement No 801413; project H-Reality.
Dave Orne of Lockheed Martin for his mentorship and enthusiasm for the paper.